\documentclass{article}
\usepackage{amsmath,graphicx}
\usepackage{caption}
\usepackage{subcaption}
\usepackage[utf8]{inputenc} 
\usepackage[T1]{fontenc}    
\usepackage{hyperref}       
\usepackage{url}            
\usepackage{booktabs}       
\usepackage{amsfonts}       
\usepackage{nicefrac}       
\usepackage{microtype}      
\usepackage{xcolor}         
\usepackage{algorithm}
\usepackage{multirow}
\usepackage{makecell}
\usepackage[hang,flushmargin]{footmisc}
\usepackage{algorithm}
\usepackage[noend]{algpseudocode}
\usepackage{pifont}
\usepackage[title]{appendix}

    \usepackage[preprint]{neurips_2022}

\title{Improving fairness for spoken language understanding in atypical speech with Text-to-Speech}

\author{%
  Helin~Wang \\
  Johns Hopkins University\\
  \texttt{hwang258@jhu.edu} \\
  \And
    Venkatesh~Ravichandran \\
  Amazon\\
  \texttt{veravic@amazon.com} \\
    \And
    Milind~Rao \\
  Amazon\\
  \texttt{milinrao@amazon.com} \\
    \And
    Becky~Lammers \\
  Johns Hopkins University School of Medicine\\
  \texttt{blammer2@jh.edu} \\
    \And
    Myra~J.~Sydnor \\
  Johns Hopkins University School of Medicine\\
  \texttt{msydnor3@jhmi.edu} \\
  \And
    Nicholas~Maragakis  \\
  Johns Hopkins University School of Medicine\\
  \texttt{nmaragak@jhmi.edu} \\
    \And
    Ankur~A.~Butala \\
  Johns Hopkins University School of Medicine\\
  \texttt{Ankur.Butala@jhmi.edu} \\
    \And
    Jayne~Zhang \\
  Johns Hopkins University School of Medicine\\
  \texttt{jz@jhmi.edu} \\
    \And
    Lora~Clawson \\
  Johns Hopkins University School of Medicine\\
  \texttt{lclawson@jhmi.edu} \\
    \And
    Victoria~Chovaz \\
  Johns Hopkins University School of Nursing\\
  \texttt{vchovaz1@jhmi.edu} \\
    \And
    Laureano~Moro-Velázquez \\
  Johns Hopkins University\\
  \texttt{laureano@jhu.edu} \\
}

\begin{document}

\maketitle

\begin{abstract}
Spoken language understanding (SLU) systems often exhibit suboptimal performance in processing atypical speech, typically caused by neurological conditions and motor impairments.
Recent advancements in Text-to-Speech (TTS) synthesis-based augmentation for more fair SLU have struggled to accurately capture the unique vocal characteristics of atypical speakers, largely due to insufficient data.
To address this issue,
we present a novel data augmentation method for \textbf{aty}pical speakers by finetuning a \textbf{TTS} model, called Aty-TTS.
Aty-TTS models speaker and atypical characteristics via knowledge transferring from a voice conversion model. Then, we use the augmented data to train SLU models adapted to atypical speech.
To train these data augmentation models and evaluate the resulting SLU systems, we have collected a new atypical speech dataset containing intent annotation.
Both objective and subjective assessments validate that Aty-TTS is capable of generating high-quality atypical speech. Furthermore, it serves as an effective data augmentation strategy, contributing to more fair SLU systems that can better accommodate individuals with atypical speech patterns.
\end{abstract}

\section{Introduction}
\label{sec:intro}

Atypical speech refers to speech patterns that deviate from typical development or the commonly accepted speaking norms for a particular age, region, or culture. 
Common speech applications, like automatic speech recognition (ASR) and spoken language understanding (SLU), often exhibit suboptimal performance when it comes to processing atypical speech \cite{moro2019study,shahamiri2021speech,DBLP:conf/interspeech/WangBh21}.
To address these performance limitations, two prevalent strategies have gained attention: transfer learning \cite{9054694} and data augmentation \cite{geng2022investigation, matsuzaka2022data}.
This paper specifically concentrates on the application of data augmentation through Text-to-Speech (TTS) synthesis of atypical speech.

TTS for atypical speakers is a largely unexplored frontier.
Soleymanpour \textit{et al.} \cite{soleymanpour2022synthesizing}
fine-tuned a TTS model that was initially trained on typical speakers, and added explicit mechanisms for dysarthria severity levels and pause insertion.
However, their approach was limited to four coarse-grained dysarthria severity levels, making it less generalizable to other datasets. 
Matsuzaka \textit{et al.} \cite{matsuzaka2022data} and Zhao \textit{et al.} \cite{zhao2021personalizing} generated dysarthric speech by chaining TTS models with Voice Conversion (VC) models,
but the speech and pause rates of the target speakers were determined by those of typical speakers, which does not always yield realistic atypical speech.
Moreover, previous research efforts have commonly evaluated ASR performance using restricted datasets including UASpeech \cite{kim2008dysarthric}, which only contains isolated words; Torgo \cite{rudzicz2012torgo}, which only features 8 dysarthric speakers; or Euphonia \cite{DBLP:conf/interspeech/MacDonaldJCHCSL21}, which is nonpublic. 
Notably, none of these studies have assessed the impact of TTS-based data augmentation on SLU performance for atypical speech or conducted subjective evaluations by experts that could indicate the quality of the synthesized speech. While some studies have employed VC techniques to generate atypical speech \cite{wang2023duta, wang2020end, huang2022towards}, 
TTS allows for the modeling of pause, speech rates, and phoneme duration for a particular target speaker or group, without the necessity for source audio \cite{ren2019fastspeech}. This capability enables the convenient creation of synthetic atypical speech tailored to specialized domains, such as complex SLU scenarios with a large number of intents, thereby improving the fairness of SLU systems.

In this paper, we present the following contributions:
(1) introduce a novel dataset of atypical speech, HeyJay, specifically curated for SLU.
(2) propose a data augmentation method to enhance the modeling of speaker characteristics and atypical characteristics for TTS.
(3) evaluate the new method in two SLU scenarios with atypical speech. To our knowledge, this is the first study that evaluates atypical speech with Fluent Speech Commands (FSC) \cite{lugosch2019speech} and SLURP \cite{bastianelli-etal-2020-slurp} intents.
(4) conduct subjective evaluations by two expert speech and language pathologists to ascertain whether the synthesized atypical speech retains the distinctive traits of the original speech it aims to emulate.\footnote{
This research was funded by the JHU + Amazon AI2AI Initiative for Interactive AI.
The source code and models are available at \href{https://github.com/WangHelin1997/Aty-TTS}{https://github.com/WangHelin1997/Aty-TTS}. The perceptual evaluations of HeyJay speakers and audio demos are shared to allow further studies: 
\href{https://wanghelin1997.github.io/Aty-TTS-Demo/}{https://wanghelin1997.github.io/Aty-TTS-Demo/}.
}

A diffusion-based TTS model \cite{popov2021grad} is adapted to atypical speech in this work, although this method can be easily transferred to other TTS models. The idea of the adaptation, called Aty-TTS, is to transfer the knowledge from an already validated VC model for atypical speech to a TTS model. 
We generate typical-atypical paired data with VC and force the decoder of the TTS model to accomplish auxiliary VC tasks from typical speech to atypical speech.
The results on HeyJay show that Aty-TTS mimics characteristics of atypical speech and can significantly improve atypical SLU.
This improvement is more pronounced when Aty-TTS is combined with other data augmentation techniques \cite{park2019specaugment,9383605}.

\section{Materials and methods}
\label{sec:method}

\subsection{New HeyJay corpus and other materials}
HeyJay is a new ongoing corpus containing atypical speech. The recorded sentences start with the wakeword "Hey, Jay", which gives the name to the corpus.
The goal of HeyJay is to provide the scientific community and software developers with a new corpus of atypical speech, including annotated transcriptions and intent to enable further research and more fair, accessible, and robust spoken-language technologies.
We use HermeSpeech \cite{park2023hermespeech} to collect atypical speech recordings at hospitals or the participant's homes. 
The data collection was approved by an Institutional Review Board, and all participants signed an informed consent and were compensated for their collaboration.
All the speakers have dysarthria, caused by neurological conditions or cerebrovascular accidents, including Parkinson's Disease, Spinocerebellar Ataxias, Amyotrophic Lateral Sclerosis, and Stroke. Dysarthria is characterized by the misarticulation of phonemes, slow (hypokinetic) or fast (hyperkinetic) speech rate, monotonous intonation, and dysphonia as main signs.
In this study, we employ 17 speakers who read sentences with the same intent as the FSC \cite{lugosch2019speech} (HeyJay-FSC partition) and 8 speakers reading sentences extracted from the SLURP \cite{bastianelli-etal-2020-slurp} dataset (HeyJay-SLURP partition), which is notable for its greater number of entities.
The recordings from FSC and SLURP were employed in this study to train the  SLU models, along with HeyJay and synthesized atypical speech. Data statistics of FSC, SLURP and HeyJay are shown in Table~\ref{tab:statistics}. 
In this study, we also used LJSpeech \cite{ljspeech17}, a dataset comprising roughly 24 hours of recordings from a single speaker, and LibriTTS \cite{zen2019libritts}, which offers 586 hours of audio from 2,456 speakers, to pre-train the TTS and VC models, respectively.

\begin{table}[t]
  \caption{Data statistics of SLU datasets. Entities are expressions that refer to objects in SLU.}
  \label{tab:statistics}
  \footnotesize
  \centering
  \begin{tabular}{ccccc}
    \hline
    &FSC \cite{lugosch2019speech}&SLURP \cite{bastianelli-etal-2020-slurp}&HeyJay-FSC&HeyJay-SLURP\\
    \hline
    Speakers & 97 & 211 & 17 & 8 \\
    Audio files & 30,043 & 141,530 & 3,765 & 1,922\\
    Duration [h] & 19 & 101.5 & 7.1& 4.2\\
    Average length [s] & 2.3 & 2.9 & 6.8 & 8.0\\
    Total Entities & 334 & 16,792 & 334 & 1,579\\
    \hline
  \end{tabular}
  \vspace{-3mm}
\end{table}
\subsection{Baseline methods}
\begin{figure}[t]
    \begin{subfigure}[c]{0.3\textwidth}
    \centering
        \includegraphics[height=2.2in]{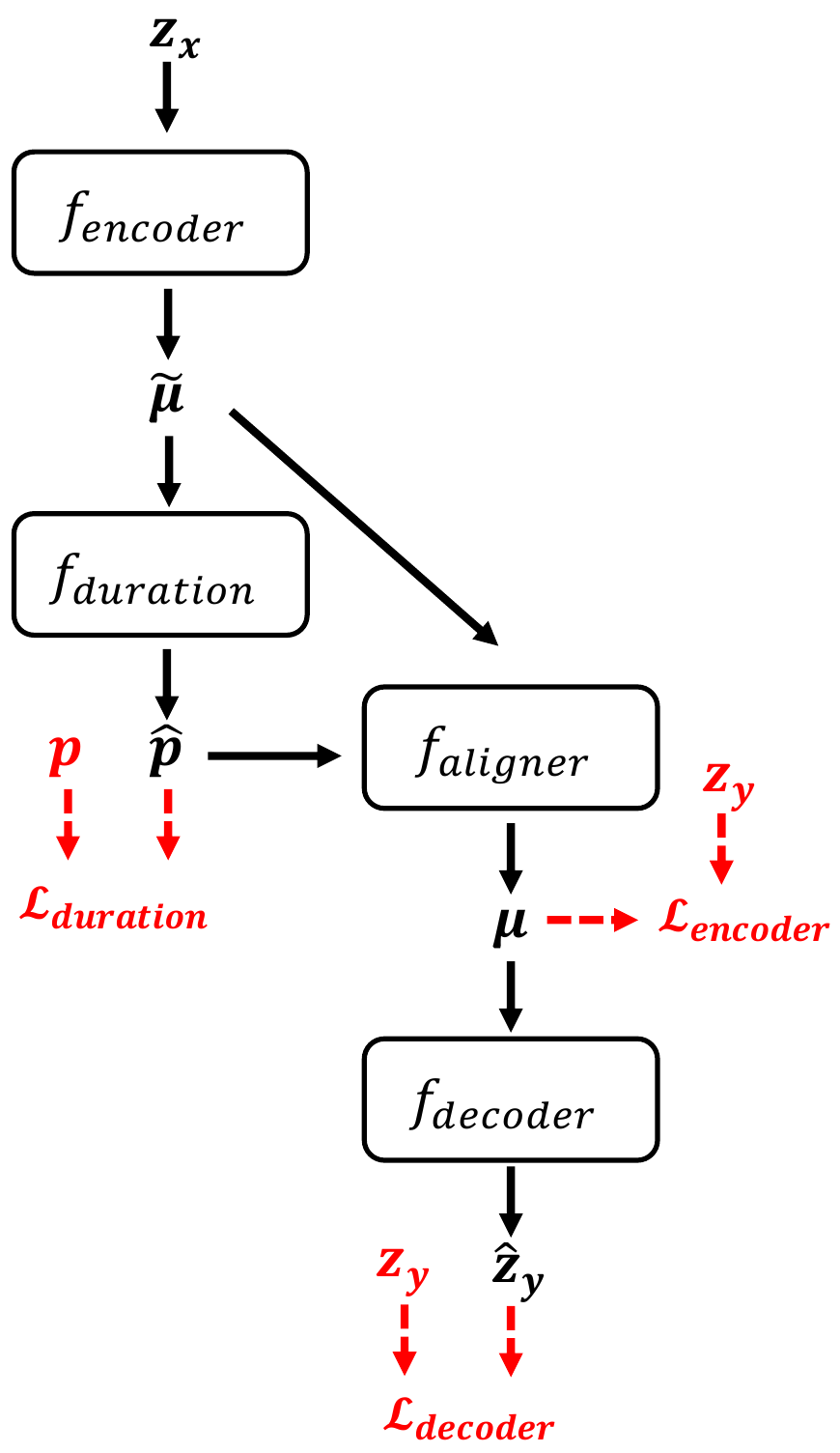}
        \caption{Grad-TTS}
        \label{fig:gradtts}
    \end{subfigure}
    \begin{subfigure}[c]{0.3\textwidth}
    \centering
        \includegraphics[height=1.2in]{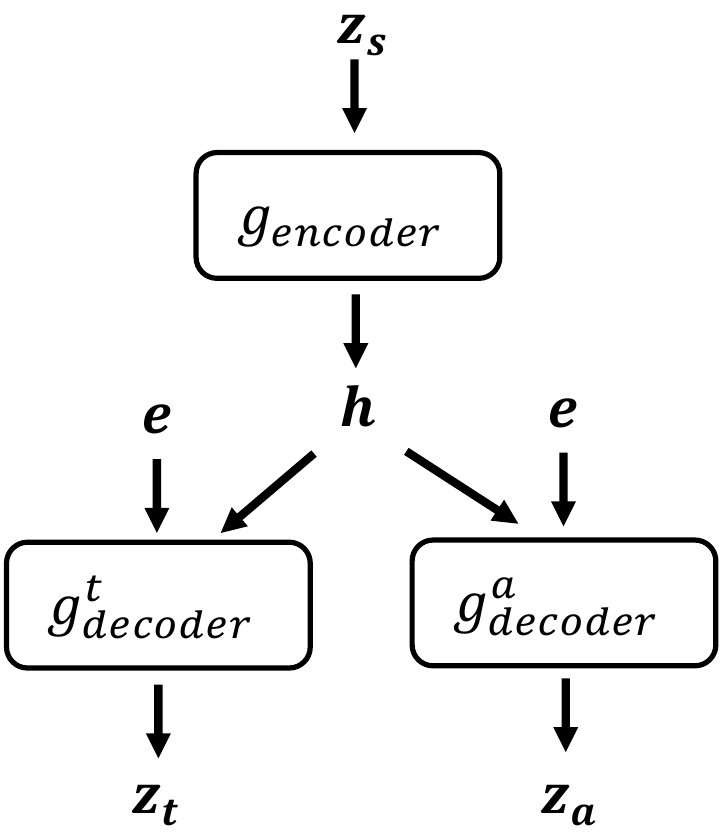}
        \caption{DuTa-VC}
        \label{fig:dutavc}
    \end{subfigure}
    \begin{subfigure}[c]{0.323\textwidth}
    \centering
        \includegraphics[height=2.2in]{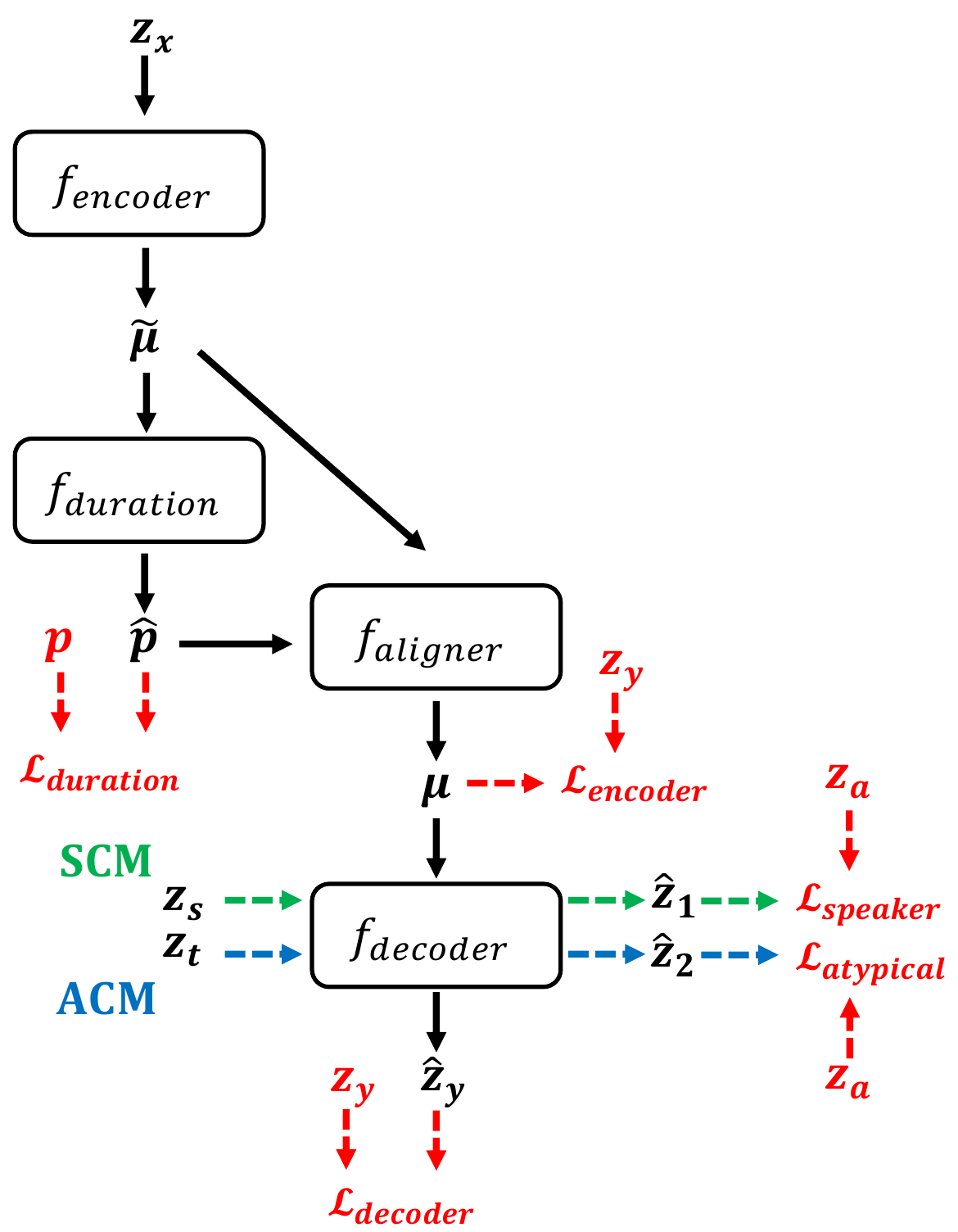}
        \caption{Aty-TTS}
        \label{fig:atytts}
    \end{subfigure}
    \caption{Training scheme of Grad-TTS (a) and Aty-TTS (c), and inference scheme of DuTa-VC (b). 
    }
    \vspace{-2mm}
\end{figure}

\textbf{Grad-TTS:} Grad-TTS \cite{popov2021grad}, whose training scheme is shown in Fig.~\ref{fig:gradtts}, is used as our baseline model.
Given an input text sequence $\boldsymbol{z_x} \in \mathbf{R}^L$ (phoneme sequence in this paper),
the model aims at generating a mel-spectrogram $\boldsymbol{z_y} \in \mathbf{R}^{T \times F}$ where $T$ is the number of acoustic frames and $F$ is the number of mel bins.
The encoder $f_{encoder}$ maps $\boldsymbol{z_x}$ into a latent feature sequence $\boldsymbol{\tilde{\mu}} \in \mathbf{R}^{L \times F}$,
followed by a duration predictor, $f_{duration}$,
which predicts phoneme durations \(\boldsymbol{\hat{p}} \in \mathbf{R}^L\). Note that $\boldsymbol{\tilde{\mu}}$ contains one latent feature representation per each phoneme contained in the input $\boldsymbol{z_x}$.
An aligner $f_{aligner}$ is then used to transform $\boldsymbol{\tilde{\mu}}$ into a latent mel-spectrogram $\boldsymbol{\mu} \in \mathbf{R}^{T \times F}$. 
For the $i$-th phoneme in  $\boldsymbol{z_x}$, the aligner expands the corresponding latent feature $\tilde{\mu}_i$ for $\lceil \hat{p}_i \rceil$ times. During training, the hard monotonic alignment \cite{kim2020glow} is used to ensure that the total length of $\boldsymbol{\mu}$ matches $T$, that is imposed by the desired output spectrogram $\boldsymbol{z_y}$.
\begin{align}
    \boldsymbol{\tilde{\mu}} &= f_{encoder}(\boldsymbol{z_x} ; \theta_{encoder}) \\
    \boldsymbol{\hat{p}} &= f_{duration}(\boldsymbol{\tilde{\mu}}; \theta_{duration}) \\
    \boldsymbol{\mu} &= f_{aligner}(\boldsymbol{\tilde{\mu}}; \boldsymbol{\hat{p}})
\end{align}
Here, $\theta_{encoder}$ and $\theta_{duration}$ are trainable parameters of the encoder and duration predictor. 
A decoder $f_{decoder}$ is then used to refine the latent mel-spectrogram and estimate the target mel-spectrogram. 
\begin{align}
    \boldsymbol{\hat{z}_y} &= f_{decoder}(\boldsymbol{\mu} ; \theta_{decoder})
\end{align}
where $\boldsymbol{\hat{z}_y} \in \mathbf{R}^{T \times F}$ is the estimated mel-spectrogram and $\theta_{decoder}$ are trainable parameters of the decoder.
There are three loss functions in Grad-TTS: encoder loss, duration predictor loss and decoder loss.
The encoder loss is applied to minimize the distance between the aligned encoder output $\boldsymbol{\mu}$ and target mel-spectrogram $\boldsymbol{z_y}$. The duration predictor loss is applied to minimize the distance between the predicted duration of the phonemes with the real phoneme duration $\boldsymbol{p} \in \mathbf{R}^L$ in the logarithmic domain. The decoder loss is used to minimize the difference between the decoder output $\boldsymbol{\hat{z}_y}$ and target mel-spectrogram $\boldsymbol{z_y}$. The mean square error (MSE) criterion is applied to all of them.
\begin{align}
    \mathcal{L}_{encoder} &= \frac{1}{T \times F}\|\boldsymbol{\mu}-\boldsymbol{z_y}\|_2^2 \\
    \mathcal{L}_{duration} &= \frac{1}{L}\|\boldsymbol{\hat{p}}-\boldsymbol{p}\|_2^2 \\
    \mathcal{L}_{decoder} &= \frac{1}{T \times F}\|\boldsymbol{\hat{z}_y}-\boldsymbol{z_y}\|_2^2
\end{align}

\textbf{DuTa-VC:} DuTa-VC is a method validated in \cite{wang2023duta} that provides successful voice conversion from typical to atypical speech.
As shown in Fig.~\ref{fig:dutavc}, for an input source typical mel-spectrogram $\boldsymbol{z}_s \in \mathbf{R}^{T^{\prime} \times F}$,
the output depends on the decoder, and can be either a typical mel-spectrogram with target speaker timbre $\boldsymbol{z_t} \in \mathbf{R}^{T^{\prime} \times F}$ via typical-to-typical VC or an atypical mel-spectrogram with the same target speaker timbre $\boldsymbol{z_a} \in \mathbf{R}^{T^{\prime} \times F}$ via typical-to-atypical VC, where $T^{\prime}$ denotes the number of acoustic frames. They share the same encoder $g_{encoder}$, where
the difference is using a decoder $g^t_{decoder}$ trained with typical speech data or using a decoder $g^a_{decoder}$ trained with atypical speech data. Both decoders have the same input, \textit{i.e.} the hidden representation $\boldsymbol{h}$ from the encoder and a speaker embedding $\boldsymbol{e}$ obtained with a model pre-trained for speaker recognition \cite{jia2018transfer}. 
In our experiments,
$g_{encoder}$, $g^t_{decoder}$ and $g^a_{decoder}$ are trained on LibriTTS.
$g^a_{decoder}$ is further finetuned on each speaker of HeyJay separately.
We do not change phoneme duration like \cite{wang2023duta} for simplification.

\subsection{Aty-TTS}
Trying to fine-tune Grad-TTS with atypical speech, speaker and atypical characteristics can be challenging to model, especially for the decoder, as the amount of atypical audio data is often small.
In our initial experiments, we found that the synthesized speech mismatched articulation characteristics with the real atypical speech, and the timbre of the synthesized speaker was often not like the target speaker. 
To overcome these problems, we propose a method to finetune the TTS model, which transfers the knowledge from a VC model to the TTS decoder.
It was validated in \cite{huybrechts2021low} that voice-converted data more closely matches the target recordings than any other auxiliary data, even when such data originates from a different style or speaker. 
As a result, the TTS decoder benefits from a broader training set with a consistent distribution.

Consequently, we enhance the training of the TTS decoder by two strategies:
speaker characteristics modeling (SCM) and atypical characteristics modeling (ACM), which are shown in Fig.~\ref{fig:atytts}.
The idea is to generate typical-atypical paired data with VC and force the TTS model to accomplish auxiliary VC tasks by adapting it with that paired data.
More specifically,
SCM guides the TTS decoder to convert the source typical mel-spectrogram $\boldsymbol{z_s}$ into the atypical mel-spectrogram with target speaker timbre $\boldsymbol{z_a}$ while ACM guides the TTS decoder to convert the typical mel-spectrogram with target speaker timbre $\boldsymbol{z_t}$ into the atypical mel-spectrogram with target speaker timbre $\boldsymbol{z_a}$.
In this way, we can use much more data from VC to train the TTS decoder.
While SCM captures both timbre and articulation, we introduced ACM specifically to improve articulation modeling, a crucial element in atypical speech. 
Our initial experiments indicate that ACM+SCM enhances TTS performance for atypical speakers more than only SCM.
The estimated mel-spectrograms $\boldsymbol{\hat{z}_1} \in \mathbf{R}^{T^{\prime} \times F}$ with SCM and $\boldsymbol{\hat{z}_2} \in \mathbf{R}^{T^{\prime} \times F}$ with ACM are obtained by
\begin{align}
    \boldsymbol{\hat{z}_1} &= f_{decoder}(\boldsymbol{z_s} ; \theta_{decoder}) \\
    \boldsymbol{\hat{z}_2} &= f_{decoder}(\boldsymbol{z_t} ; \theta_{decoder})
\end{align}
The MSE loss is applied to minimize the difference between the decoder outputs and the atypical mel-spectrogram with target speaker timbre $\boldsymbol{z_a}$.
\begin{align}
    \mathcal{L}_{speaker} &= \frac{1}{T^{\prime} \times F}\|\boldsymbol{\hat{z}_1}-\boldsymbol{z_a}\|_2^2 \\
    \mathcal{L}_{atypical} &= \frac{1}{T^{\prime} \times F}\|\boldsymbol{\hat{z}_2}-\boldsymbol{z_a}\|_2^2
\end{align}
Aty-TTS is trained with all the loss functions:
\begin{align}
    \mathcal{L} &= \mathcal{L}_{encoder} + \mathcal{L}_{duration} + \mathcal{L}_{decoder} + \mathcal{L}_{speaker} + \mathcal{L}_{atypical}
\end{align}
We use LibriTTS data as the source audio for DuTa-VC.
Aty-TTS models are pre-trained on LJSpeech and then finetuned on HeyJay for each atypical speaker separately.
After generating the mel-spectrogram with the decoder, 
the pre-trained HiFi-GAN vocoder \cite{kong2020hifi} trained on LJSpeech is used to reconstruct the audio waveform. We finetune the vocoder for each atypical speaker separately.

\section{Experiments}
\label{sec:experiments}
In this study, we compare a new data augmentation method that generates synthetic speech, Aty-TTS, with other baseline methods, including WaveAug \cite{9383605}, SpecAug \cite{park2019specaugment}, Grad-TTS \cite{popov2021grad}, DuTa-VC \cite{wang2023duta},. This was done by comparing the performance of different atypical SLU systems trained with these augmented data. Moreover, we further investigate the impact of varying the quantity of synthesized data on SLU training.
\subsection{Experiment Setups}

Aty-TTS has the same architecture as Grad-TTS \cite{popov2021grad}, and the VC model has the same architecture as DuTa-VC \cite{wang2023duta}.
Following their settings, we use 80-dimensional mel-spectrograms with Short-Time Fourier Transform (STFT) window size of 46.4 ms and hop size of 11.6 ms.
All the audio files are re-sampled at 22.05 kHz.
The number of frames for the Aty-TTS is 172 at training (2 seconds).
Adam optimizer is employed with initial learning rates $1 \times 10^{-4}$ for pre-training and $5 \times 10^{-5}$ for finetuning, respectively.
Batch sizes are set to 64 and 32 with 200 epochs and 50 epochs, respectively. We use 20 hours of VC augmentation for each atypical speaker in Aty-TTS.

\begin{table}[t]
  \caption{Intent Classification Accuracy (ICA) results on HeyJay-FSC and FSC.}
  \label{tab:fsc}
  \footnotesize
  \centering
  \begin{tabular}{c|c|c|c|c}
    \hline
    \multirow{2}{*}{Method} & \multicolumn{3}{c|}{HeyJay-FSC} & \multirow{2}{*}{FSC} \\
    \cline{2-4}
     &Low&High&All&\\
    \hline
    \multicolumn{1}{l|}{HuBERT \cite{hsu2021hubert}} & 90.5 & 77.0 &  85.4& 98.7\\
    \multicolumn{1}{l|}{+WaveAug \cite{9383605} + SpecAug \cite{park2019specaugment}} & 94.0 & 82.4 & 90.8 & 99.0\\
    \multicolumn{1}{l|}{+WaveAug \cite{9383605} + SpecAug \cite{park2019specaugment} + Grad-TTS \cite{popov2021grad}} & 95.4& 85.1 & 91.9 & 99.4\\
    \multicolumn{1}{l|}{+WaveAug \cite{9383605} + SpecAug \cite{park2019specaugment} + DuTa-VC \cite{wang2023duta}}  &  \textbf{96.1} &  89.6& 93.6 & 99.5\\
    \multicolumn{1}{l|}{+WaveAug \cite{9383605} + SpecAug \cite{park2019specaugment} + \textbf{Aty-TTS (ours)}}  & 95.9 & \textbf{90.2} & \textbf{93.7} & \textbf{99.5}\\
    \hline
  \end{tabular}
  \vspace{-4mm}
\end{table}

\begin{table}[t]
  \caption{SLU F1 results on HeyJay-SLURP and SLURP.}
  \label{tab:slurp}
  \footnotesize
  \centering
  \begin{tabular}{c|c|c|c|c}
    \hline
    \multirow{2}{*}{Method} & \multicolumn{3}{c|}{HeyJay-SLURP} & \multirow{2}{*}{SLURP} \\
    \cline{2-4}
     &Low&High&All&\\
    \hline
    \multicolumn{1}{l|}{HuBERT \cite{hsu2021hubert}} & 79.79 & 68.53 & 73.69 & 75.72\\
    \multicolumn{1}{l|}{+ WaveAug \cite{9383605} + SpecAug \cite{park2019specaugment}} & 80.54 & 72.38 & 76.13 & 77.25\\
    \multicolumn{1}{l|}{+ WaveAug \cite{9383605} + SpecAug \cite{park2019specaugment} + Grad-TTS \cite{popov2021grad}} & 80.71 & 73.55 & 77.15& 78.60\\
    \multicolumn{1}{l|}{+ WaveAug \cite{9383605} + SpecAug \cite{park2019specaugment} + DuTa-VC \cite{wang2023duta}}  & 79.03  & 67.50 & 73.10 & 73.44\\
    \multicolumn{1}{l|}{+ WaveAug \cite{9383605} + SpecAug \cite{park2019specaugment} + \textbf{Aty-TTS (ours)}}  & \textbf{80.87} & \textbf{75.96} & \textbf{78.06} & \textbf{79.02}\\
    \hline
  \end{tabular}
  \vspace{-4mm}
\end{table}

\subsection{SLU and objective evaluation metrics}
\textbf{SLU with FSC:} We used FSC \cite{lugosch2019speech} training set and HeyJay-FSC to finetune a HuBERT base model \cite{hsu2021hubert}, and then test on FSC test set and HeyJay-FSC. 
Speakers from HeyJay-FSC were categorized into two groups: (i) \textbf{Low}-dysarthria (11 speakers) and (ii) \textbf{High}-dysarthria (6 speakers)\footnote{We considered that participants with a dysarthria severity equal or lower than 1.5, in a scale from 0 to 4, had Low-dysarthria severity. Participants with scores higher than 1.5
had High-dysarthria severity.
}.
We divided the 17 speakers randomly into five distinct parts and employed 5-fold cross-validation to conduct leave-speakers-out experiments, where speakers for test were not seen during training.
20 hours of synthesized atypical speech were used, which is similar to the duration of FSC.
Following other experiments on FSC \cite{lugosch2020using, yang2021superb, yang21c_interspeech}, intent classification accuracy (ICA) was calculated.

\noindent \textbf{SLU with SLURP:} We used SLURP \cite{bastianelli-etal-2020-slurp} training set and synthesized data with Aty-TTS by all speakers in HeyJay-FSC to finetune a HuBERT base model \cite{hsu2021hubert}, and then test on SLURP test set and HeyJay-SLURP. 
We excluded far-range audios from the SLURP test set, as our focus in this work is not on noisy speech.
None of the speakers or sentences included in the test set were exposed to the model during the training phase.
Test speakers from HeyJay-SLURP were categorized into two groups: (i) \textbf{Low}-dysarthria and (ii) \textbf{High}-dysarthria, each group containing 4 speakers.
We used 100 hours atypical speech synthesized with Aty-TTS to match the duration of SLURP.
Following previous studies on SLURP \cite{arora2022espnet, peng2022branchformer, speechbrain}, SLU F1 \cite{bastianelli-etal-2020-slurp} was calculated.

\subsection{Subjective Evaluation Metrics}
Two speech and language pathologists with more than 10 years of experience blindly evaluated both the real and generated speech samples generated by Aty-TTS. Each pathologist was provided with eight audio clips for each speaker and was asked to assess the overall degree of atypical speech traits. The assessment scores spanned from 0 to 4, where 0 indicated no abnormality and 4 indicated a severe condition, following the Rating Scale for Deviant Speech Characteristics protocol described in chapter three of~\cite{duffy2012motor}. For each set of real and synthetic speech, we computed the average score for the eight clips from each of the two pathologists. Statistical measures were calculated to compare the differences between the scores of the real and generated speech for each particular trait: mean absolute error (MAE) which provides a mean value representation of the error, root-mean-square error (RMSE) which offers insight into the magnitude of errors and penalizing larger mistakes, and the coefficient of determination (R$^2$) which reveals the proportion of data variance.


\subsection{Results}


The SLU performance for the FSC and HeyJay-FSC datasets is detailed in Table~\ref{tab:fsc}. To maintain consistency across experiments, we keep factors such as the number of training iterations and the amount of synthesized data constant. Specifically, we finetune a pre-trained Grad-TTS model for each speaker in the HeyJay-FSC subset.
Our findings indicate that performing SLU on HeyJay-FSC is considerably more challenging than on the FSC dataset. The impact of high-dysarthria speech on SLU performance is also more pronounced compared to low-dysarthria speech. Signal-based data augmentation techniques (WaveAug and SpecAug) significantly enhance intent classification. Methods relying on synthesized speech further improve performance. Among these, Grad-TTS performs the least effectively. In contrast, both Aty-TTS and DuTa-VC successfully model atypical characteristics, yielding notably better results, especially for speakers with high levels of dysarthria.
Table~\ref{tab:slurp} shows the SLU performance for the SLURP and HeyJay-SLURP datasets.
Aty-TTS outperforms all other techniques across both dysarthria groups, demonstrating particularly marked improvements for speakers with high levels of dysarthria compared to Grad-TTS. DuTa-VC's performance is suboptimal, due to the noisy and reverberant recordings of the SLURP, which are used as source audios in VC.
In such challenging acoustic conditions, TTS-based augmentations prove to be a more effective strategy than VC-based ones. 
While VC is also employed for augmentation in Aty-TTS training, clean source data can be used for VC, such as LibriTTS.
In summary, Aty-TTS is more versatile than VC in these scenarios, and it enhances the fairness of SLU systems by narrowing the performance gap between typical and atypical speech, as well as between speech with low and high levels of dysarthria.

\begin{figure}[t]
    \begin{subfigure}[c]{0.45\textwidth}
    \centering
        \includegraphics[height=1.3in]{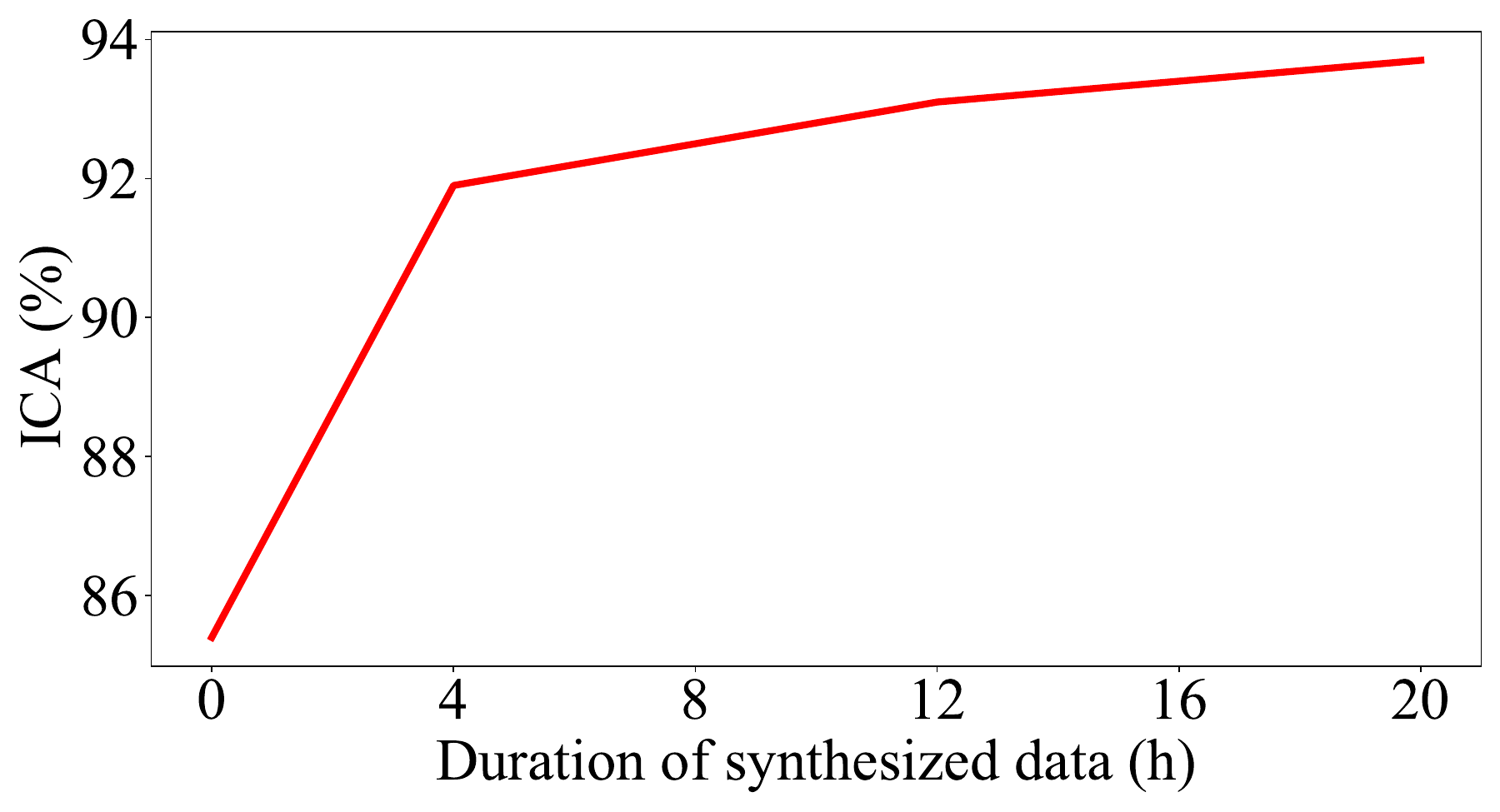}
        \caption{ICA results on HeyJay-FSC}
        \label{fig:fsc}
    \end{subfigure}
    \begin{subfigure}[c]{0.45\textwidth}
    \centering
        \includegraphics[height=1.3in]{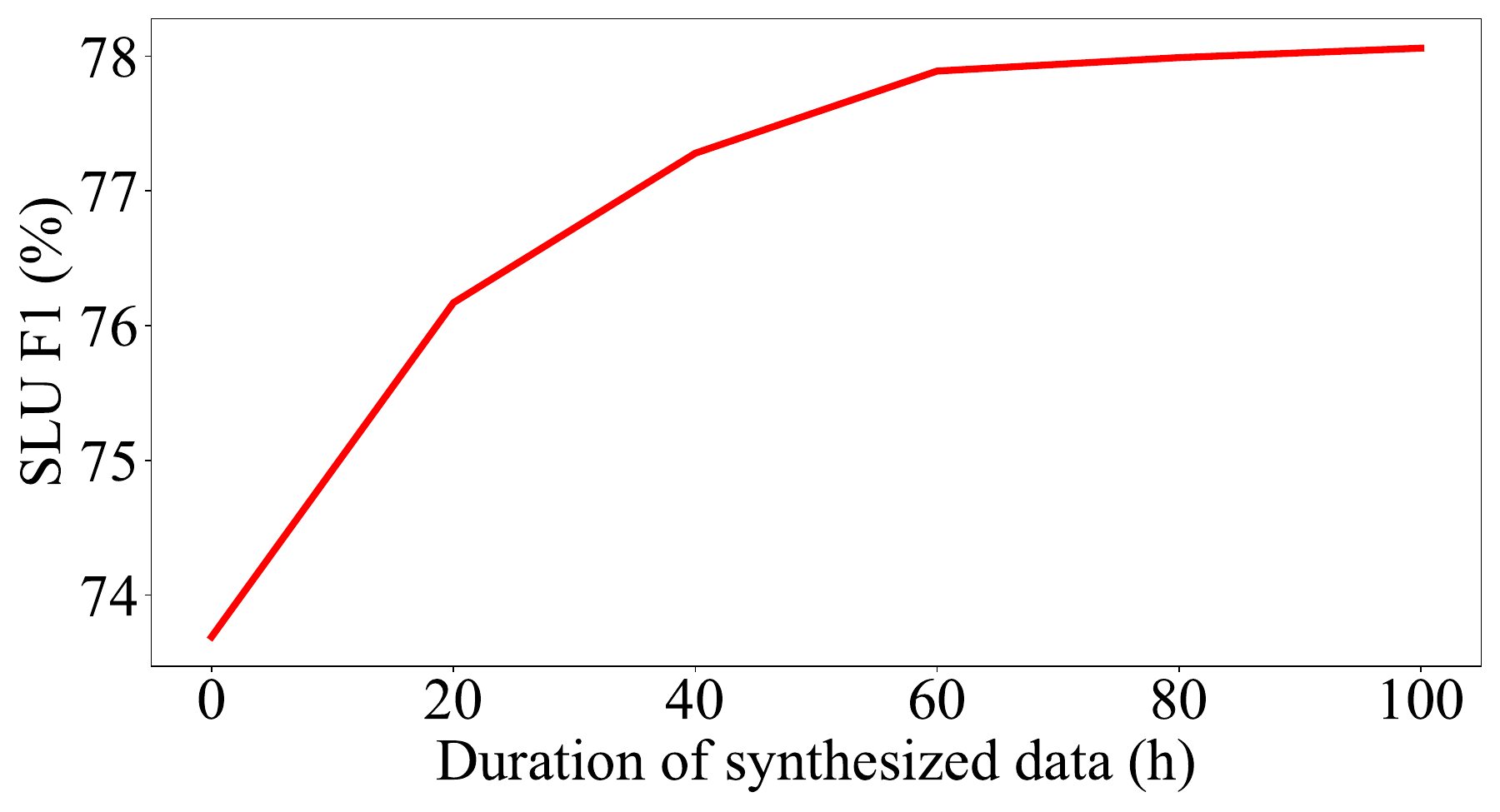}
        \caption{SLU F1 results on HeyJay-SLURP}
        \label{fig:slurp}
    \end{subfigure}
    \caption{Influence of different amounts of synthesized data on HeyJay.}
    \vspace{-4mm}
\end{figure}

Additionally, we use a range from 0 to 20 h of data synthesized by Aty-TTS for HeyJay-FSC and from 0 to 100 hours for HeyJay-SLURP. As illustrated in Fig.\ref{fig:fsc}, even a modest amount of synthesized data (e.g., 4 h) yields a substantial improvement in performance on the HeyJay-FSC. While additional data does lead to better accuracy, the performance gains plateau, showing marginal improvements beyond 8 h of synthesized data. In contrast, for HeyJay-SLURP, as depicted in Fig.\ref{fig:slurp}, performance continually improves with the inclusion of more synthesized data. This is likely due to the greater complexity of the SLURP dataset compared to FSC and a more fine-gained metric (SLU F1) is applied. An optimal performance is observed when utilizing 60 h of synthesized data.


\begin{table}[t]
  \caption{MAE, RMSE and R$^2$ between the real and synthetic speech on 12 atypical speech traits, including overall dysarthria severity (T1), overall articulation severity (T2), imprecise consonants (T3), prolonged phonemes (T4), irregular breakdowns (T5), distorted vowels (T6), overall voice quality (T7), harsh voice (T8), hoarse/wet voice (T9), breathy voice (T10), strained/strangled voice (T11), stoppages (T12).
  }
  \label{subject}
    \footnotesize
  \centering
  \begin{tabular}{c|p{4mm}p{4mm}p{4mm}p{4mm}p{4mm}p{4mm}p{4mm}p{4mm}p{4mm}p{4mm}p{4mm}p{4mm}}
    \hline
    \multirow{2}{*}{Metric} & \multicolumn{12}{c}{Trait} \\
    \cline{2-13}
    &T1& T2& T3& T4& T5& T6& T7& T8& T9& T10& T11& T12\\
    \hline
    MAE$\downarrow$ &0.60 &0.56 &0.57 & 0.29& 0.47& 0.44& 0.32& 0.41& 0.34& 0.15& 0.25& 0.01\\
    RMSE$\downarrow$& 0.75& 0.73& 0.72& 0.40& 0.58& 0.58& 0.51& 0.48& 0.50& 0.24& 0.32& 0.06\\
    R$^2$$\uparrow$  &0.56 & 0.57& 0.52& 0.73 & 0.48& 0.48& 0.66& 0.08& 0.62& 0.90& 0.82& 1.00 \\
    \hline
  \end{tabular}
  \vspace{-2mm}
\end{table}


Table~\ref{subject} presents the findings from our subjective evaluation metrics. Aty-TTS effectively models various speech characteristics, including articulatory prolonged phonemes, overall voice quality, hoarseness, wetness, breathiness, strained or strangled voice, and stoppages, with MAE below 0.34, RMSE below 0.51, and R$^2$ over 0.62.
These values indicate a strong match in terms of both dysarthria severity and articulation quality between the real and synthetic speech samples.

\section{Conclusions}
\label{sec:conclusions}
We proposed a data augmentation method to improve TTS for atypical speakers with VC.
The TTS model could generate high-quality atypical speech and be an effective data
augmentation method for more fair SLU.
In future works, we will (1) record more speakers (2) evaluate the method on ASR (3) explore multi-speaker TTS for atypical speakers.

\medskip

{
\small
\bibliographystyle{IEEE}
\bibliography{refs}
}

\end{document}